\documentclass[usenatbib]{HAWC2015}
\usepackage{aas_macros}
\usepackage{mathptmx}       
\usepackage{helvet}         
\usepackage{courier}        

\usepackage[numbers,sort,comma]{natbib}
\usepackage{graphicx}        

\include{def}
\begin{document}
\title*{Active Galactic Nuclei: The TeV Challenge}
\titlerunning{AGN: TeV Challenge}
\author{R. Blandford$^1$, W. East$^1$, K. Nalewajko$^1$, Y. Yuan$^1$ \&  J. Zrake$^1$.}
\authorrunning{Blandford} 
\institute{$^{1}$KIPAC, Stanford University, CA 94305, USA}

\maketitle
\vskip -3. cm  
\abstract{Jets associated with Active Galactic Nuclei (AGN) have been observed for almost a century, initially at optical and radio wavelengths. They are now widely accepted as "exhausts" produced electromagnetically by the central, spinning, massive black hole and its orbiting, accreting gas. Observations at X-ray and, especially, $\gamma$-ray energies have transformed our understanding of how these jets evolve dynamically, accelerate electrons (and positrons) and radiate throughout the entire electromagnetic spectrum. Some new approaches to modeling the powerful and rapidly variable TeV emission observed from many blazars are sketched. Observations at the highest TeV energies, to which the High Altitude Water Cherenkov Gamma-Ray Observatory (HAWC) will contribute, promise crucial discrimination between rival models of AGN jets.}
\vskip -0.5 cm
\setlength{\unitlength}{1.cm}
\section{Introduction}
\label{sec:Int}
Gamma-ray astronomy has opened up a huge region of the electromagnetic spectrum. From $\sim100$~keV to $\sim100$~TeV lies 30 octaves of photon energy, in contrast to the single octave associated with the visible band. Of course, there is much more physics (and chemistry) going on at $\sim1$~eV energies but this does at  least suggest that there is plenty of $\gamma$-ray discovery space to be explored, especially when one adds in the rapidly maturing observational capabilities using cosmic rays, neutrino and gravitational radiation. HAWC, which will operate from $\sim0.1-100$~TeV, has the opportunity to be a major player in high energy astrophysics. It will complement Fermi Gamma Ray Space Telescope, (henceforth Fermi)\cite{2009ApJ...697.1071A}, by operating at higher energy and the Atmospheric Cherenkov Telescopes~\cite{2009ARA&A..47..523H}, 
including especially the future Cherenkov Telescope Array (CTA),.by observing a large fraction of the sky at all times (like Fermi).

TeV $\gamma$-ray astronomy has already grown from making a handful of tentative detections to catalogs of nearly 200 sources (e.g. http://tevcat.uchicago.edu). The majority of these sources are AGN, with flux-limited samples dominated by the emission produced by relativistic jets beamed towards us. These jetted AGN are called ``Blazars'', which are characterized by identification with a compact radio source, rapid optical variability and strong polarization.  (Fermi has identified over 1000 blazars.) Blazars are often separated into ``Flat Spectrum Radio Quasars'' (FSRQ), which are higher power and distant and exhibit a thermal continuum and optical emission lines and ``BL Lac'' sources (BLL) which are lower power and closer and where emission lines are weak. The BLL are divided, spectrally,  into Low frequency peaked  (LBL) sources, with synchrotron emission peaking in the infrared and Compton emission peaking at GeV energy and High frequency peaked (HBL) sources, where the synchrotron component peaks in the X-rays and the Compton component appears at TeV energy. FSRQ are mostly spectrally similar to LBL at high energies~\cite{1995ApJ...444..567P}.
It is hard to make the taxonomy precise and definitions have shifted as observational capabilities have evolved. Selection effects are also challenging to address.

Radio observations of jets exhibit ``superluminal'' expansion -- features apparently moving across the sky faster than light -- as was inferred on the grounds that the small source sizes inferred on the basis of the observed variability would otherwise lead to catastrophic inverse Compton losses~\cite{1966Natur.211..468R}. This strongly suggests that jets are relativistic outflows and their synchrotron and Compton emission is beamed within a cone with opening angle $\sim\Gamma^{-1}$ about the velocity of the emitting plasma and $\Gamma$ is the associated bulk Lorentz factor. It is now generally accepted that the emission comes from a range of radii with the magnetic field responsible of the synchrotron emission decreasing with radius. The soft photons scattered as $\gamma$-rays mostly originate within the jet in the BL Lac objects and as thermal emission emitted externally by the accreting gas in the FSRQs. (Hadronic models of blazar jets have also been developed but seem problematic and will not be discussed further here)

Blazars account for the majority of the high energy $\gamma$-ray background above 100 GeV~\cite{2015ApJ...800L..27A}. The manifest, low pair production optical depth out to the highest redshift FSRQ constrains the infrared background to be not much more than the summed intensity from observed galaxies and stars, contrary to what was claimed by observational cosmologists~\cite{2015arXiv150204166B}.
\begin{figure}[]
          \includegraphics[scale=0.5]{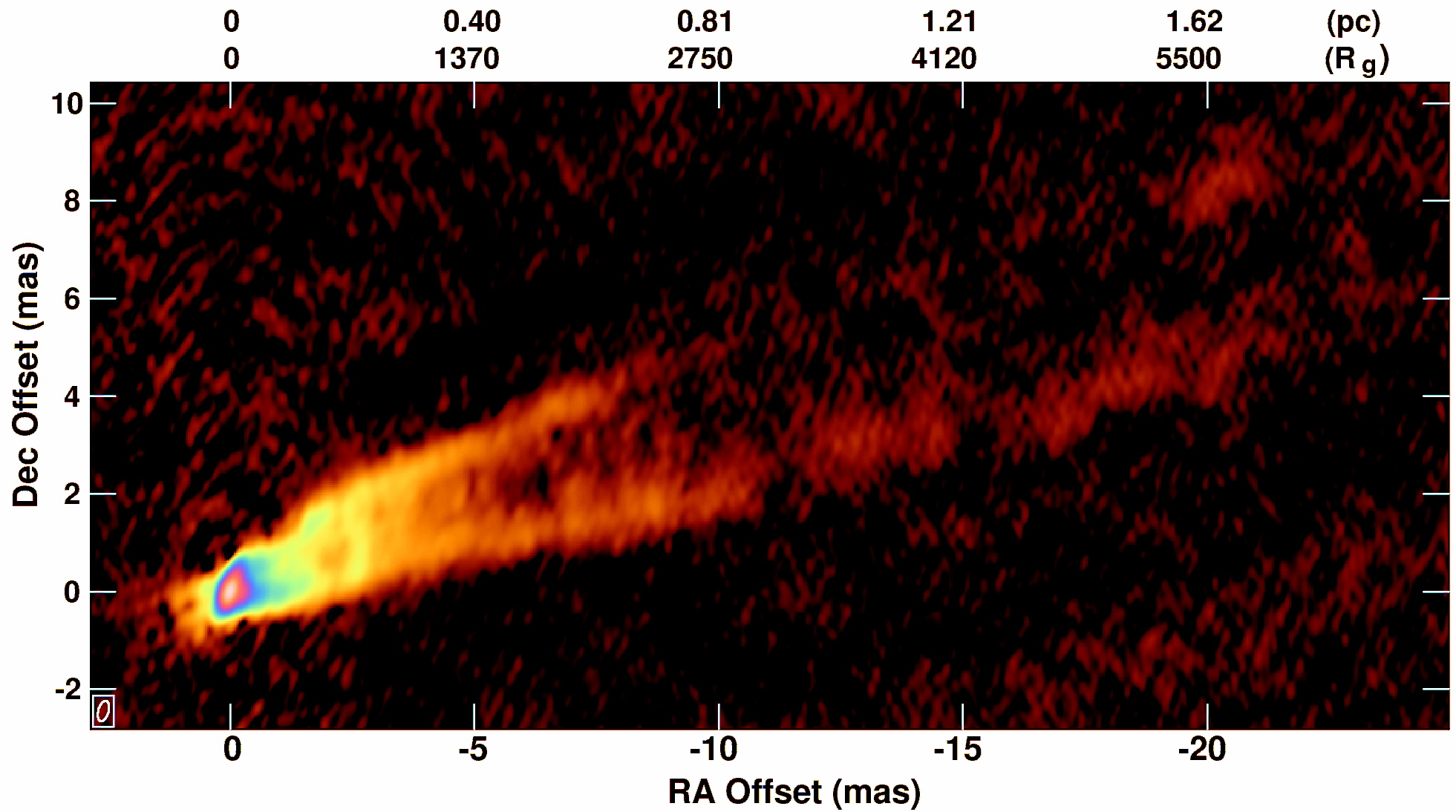}
        \caption{43~GHz radio observations of the jet in M87 made with the VLBA~\cite{2008JPhCS.131a2053W}. 1 mas is roughly equivalent to $\sim300GM/c^2$, The jet is inclined to the line of sight with an angle that has been variously estimated to lie in the range $\sim15^\circ-45^\circ$.   \label{fig:m87} }
\end{figure}

\section{Some Observations of AGN Jets}
\label{sec:agn}
Much of what is now known about Active Galactic Nuclei (AGN) and, more generally, astrophysical jets is exemplified by M87. (The M refers to the Messier catalog which was initially produced to {\it remove} objects which might interfere with the study of comets!) In 1918, Heber Curtis, working at the Lick Observatory noticed a ``curious straight ray'' emanating from the center of the M87 and this was the first extragalactic jet. Although M87 is a radio galaxy, it is not a blazar. However, it surely would be a BLL if our line of sight were oriented more closely to the direction of the jet.~\cite{1978bllo.conf..328B}. M87 is powered by accretion onto a seven billion solar mass black hole in the nucleus of a giant elliptical galaxy located in a rich cluster of galaxies~\cite{2011ApJ...729..119G}. The current accretion rate appears to be quite low and the central luminosity is extremely small compared with the Eddington limit, $L_E=4\pi GMm_pc/\sigma_T$, where $M$ is the hole mass. The jet, which is observed throughout the electromagnetic spectrum, appears to be much more powerful than the accretion disk. The low radio frequency observations (Fig.~\ref{fig:m87}) reveal that it fueled a succession of ``double'' radio sources which detached as bubbles and then floated upwards in the surrounding intracluster gas~\cite{2001ApJ...554..261C}.  At the upper end of M87's spectrum, $\sim20$~TeV gamma rays have been observed and variability measured on $\sim~1$ day timescales~\cite{2006Sci...314.1424A}. The most detail is provided by Very Long Baseline Interferometry at mm wavelengths which can resolve an edge-brightened outflow down to $\sim100$ gravitational radii and structure down to $\sim10$ gravitational radii~\cite{2012Sci...338..355D}. There is some observational evidence that the $\gamma$-ray variations are associated with the inner radio jet~\cite{2015arXiv150205177H}. M87 is a prime target for the Event Horizon Telescope which is intended to observe structure influenced by the strong gravitational fields around the event horizon using submillimeter telescopes, (including the Gran Telescopio Milim\'etrico, just uphill from HAWC), and ALMA.  Additional crucial observations are likely to come over the next few years using the Jansky Very Large Array and the Astro-H X-ray satellite.
\begin{figure}[]
          \includegraphics[scale=0.8]{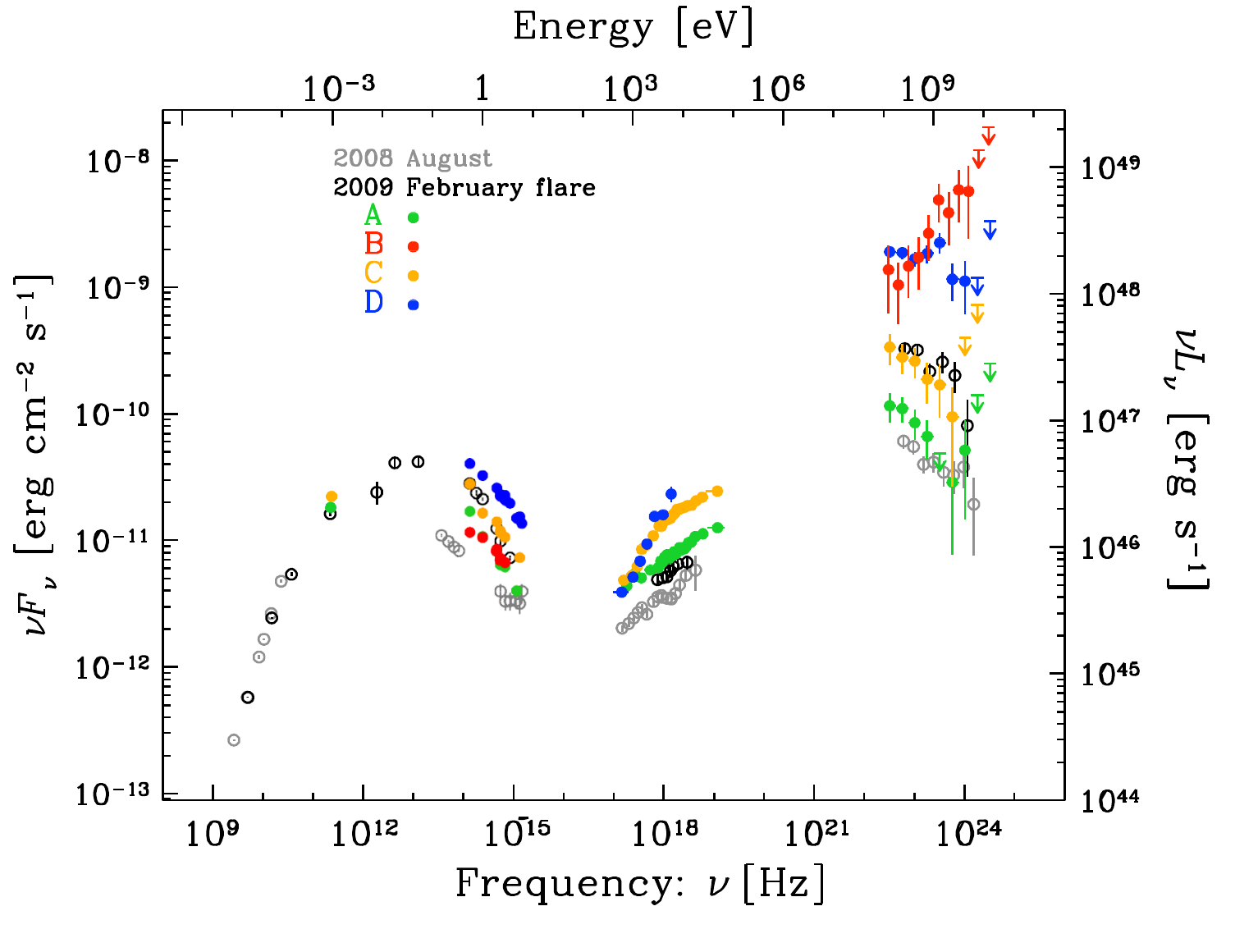}
        \caption{Recent spectral variability of the FSRQ 3C279~\cite{2015arXiv150204699H}. This exhibits the large variation and flux associated with gamma ray flares.   \label{fig:3c279} }
\end{figure}

Another prototypical jet is that associated with the first quasar, 3C 273, which was discovered in 1963. This is identified with a $\sim20$~L$^\ast$ giant elliptical galaxy. The bolometric flux from the quasar along the line of sight (dominated by ultraviolet emission) is about ten times greater than that of the galaxy. The black hole mass measurement is close to $\sim1$~billion M$_\odot$~\cite{2004ApJ...613..682P}. The associated Schwarzschild radius is $\sim3$~billion km, equivalent to a few hours. (We should not cease to be amazed that a source this small can outshine the host galaxy!) The one-sided jet has been well-mapped at radio, optical and X-ray wavelengths. The measured expansion speed of features in the jet at a distance of $\sim10$~pc from the black hole is $\sim10$~c suggesting that $\Gamma\sim10$ and that the jet is directed within $\sim6^\circ$ of the line of sight. The jet is observed out to a projected radius of $\sim70$~kpc and an actual radius of at least $\sim700$~kpc in true length protruding well beyond the galaxy. The optical emission is found to be strongly linearly polarized to a degree $\sim0.1-0.2$ in the optical band, strongly suggesting that we are observing nonthermal synchrotron radiation. This requires that relativistic electrons (and quite possibly positrons) are accelerated efficiently close to the locations where they are observed to radiate. One of the biggest debates, right now, is the site of the gamma-ray emission~\cite{2010ApJ...714L..73A}. Flares involving $\sim15$~GeV photons have been observed with timescales as short as a few hours~\cite{2013A&A...557A..71R}. However, these photons will pair produce on $\sim30$~eV ultraviolet photons and we know the luminosity of the source in this band. The radius of the sphere from which the photons can escape -- the ``gammasphere'' -- is $R\sim10f_\gamma$~ pc where $f_\gamma$ is the fraction of these photons that are emitted or (Thomson) scattered roughly perpendicular to the jet at the gammasphere. As this is in the middle of the broad emission line region, $f_\gamma$ cannot be too small. For illustration, if $f_\gamma\sim0.03$, then $R\sim1$~lt yr. Now the shortest variation time, measured at lower energy is $t_\gamma\sim1$~hr and the size of the source associated with this generally estimated to be $\sim\Gamma ct_\gamma\sim10$~lt hr. This is a small fraction of the size of the jet. Although this is not an observationally rigorous argument, it does raise the question of how are particles accelerated so efficiently within such a small portion of a large source? Similar questions can be asked of some Pulsar Wind Nebulae and Gamma Ray Bursts (GRB).

Another example is provided by the famous quasar 3C279. This has been extensively monitored at many wavelengths~\cite{2015arXiv150204699H}, (Fig.~(\ref{fig:3c279})). There are large variations with time scales as short as $\sim2$~hr which are sometimes, though not always, correlated at optical wavelengths. Optical polarization as high as $\sim0.3$ is reported with the plane of polarization seemingly swinging through angles of at least $\sim200^\circ$ suggestive of the presence of a strong magnetic field in the emission region.

\begin{figure}[t!]
          \includegraphics[width=4in]{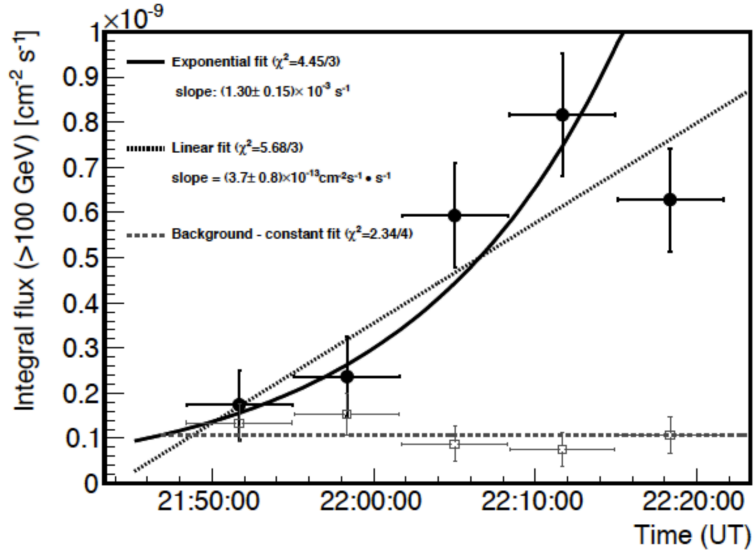}
        \caption{Rapid variation of the FSRQ PKS 1222+21~\cite{2011ApJ...730L...8A}. Flux doubling timescales of 10 min. were reported.  Even more rapid variability is reported in other sources. \label{fig:pks1222} }
\end{figure}

Even more rapid TeV variability has been reported in several sources including the FSRQ PKS 1222+21~\cite{2011ApJ...730L...8A}, (10 min., Fig.~(\ref{fig:pks1222})), the BLLs PKS 2155-304,~\cite{2010A&A...520A..83H} (2 min.) MKN 421~\cite{1996Natur.383..319G} (changing by a factor 20 in a half hour) and MKN 501~\cite{2007ApJ...669..862A} ($\sim2$~min with the high energy photons following the low energy photons with delays $\sim4$~min.).

Further evidence is provided by campaigns which cross-correlate radio and mm observations with $\gamma$-ray observations~\cite{2014MNRAS.445..428M}. The association is not strong, but when it happens the radio flare follows the $\gamma$-ray flare and presumably originates downstream from it. One well-studied example is 3C120~\cite{2015arXiv150503871C} where it has been argued that the $\gamma$-rays originate just outside the emission line cloud region at $\sim1$~lt year. $\gamma$-ray variation on a time scale of two weeks is reported, while the Lorentz factor is $\Gamma\sim6$.  

\section{Accreting Black Holes}
\label{sec:abh}
\begin{figure}[]
          \includegraphics[scale=1.1]{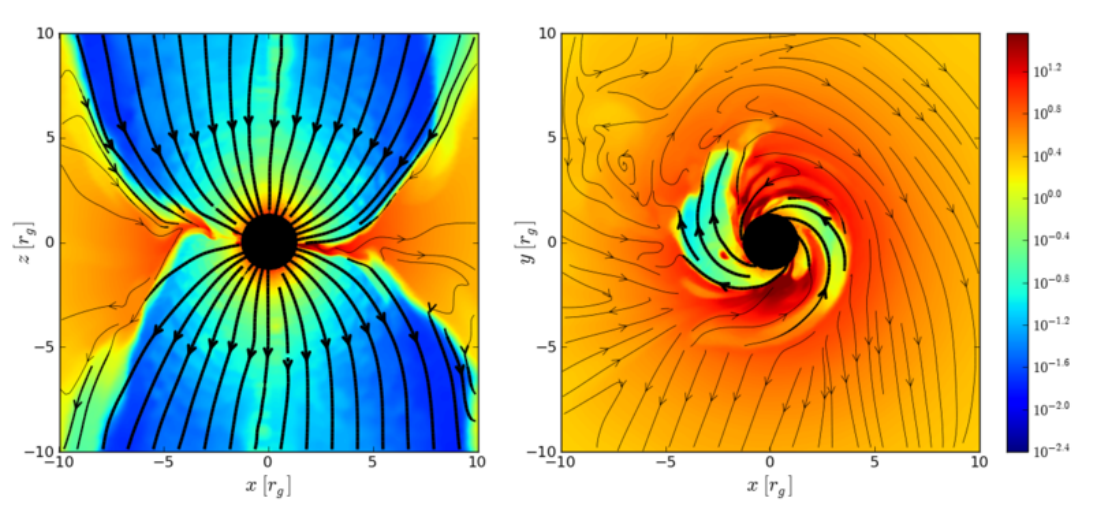}
        \caption{$3+1$D Relativistic MHD simulation of a jet formed by a spinning magnetized black hole~\cite{2012MNRAS.423.3083M}. The jets are surprisingly efficient in extracting rotational energy from the black hole and maintain their integrity in the face of large external perturbations. Interestingly, the simulations show large, quasi-periodic oscillations which might even be detectable in TeV $\gamma$-rays. \label{fig:mck} }
\end{figure}

There is now very good evidence that most normal galaxies possess a massive, nuclear black hole. The masses range from less than a million M$_\odot$ to more than $\sim10$~billion solar masses. They are believed to be the prime movers of AGN, though we are still not confident that we understand how they operate. Astrophysical black holes are described by the Kerr metric a remarkable solution of the source-free field equations of general relativity. The solution is fully parametrized by the mass $m=1.5\times10^{13}M_8{\rm cm}\equiv500M_8$~s $\equiv2\times10^{62}M_8$~erg (where $M_8=M/10^8$M$_\odot$ and $G=c=1$) and the angular momentum per unit mass, $a$, measured in the same units. The angular momentum per unit mass is limited by $-m<a<m$ and the radius and area of the event horizon are $r_+=m+(m^2-a^2)^{1/2}$, $A=8\pi mr_+$. Another critical radius is that of the so-called ergosphere, $r_e=m+(m^2-a^2\cos^2\theta)^{1/2}$, within which all particles have to rotate in a prograde sense with the hole. Now, classically, the area of the horizon cannot decrease and we can rewrite it as $A=16\pi m_0^2$, where $m_0=m[\{1+(1+a^2/m^2)^{1/2}\}/2]^1/2$ is the ``irreducible mass''. What this means is that there is rotational energy $m-m_0$ up to $0.29m$ available and extractable, in principle, by classical process. The means by which rotational energy is extracted in practice is believed to involve electromagnetic torque associated with magnetic flux that threads the event horizon and which is supported by external current presumably associated with the accretion disk. This interaction is irreversible in the sense that there is inevitable dissipation within the event horizon which increases $m_0$. The hole's spin twists and collimates the magnetic flux that threads the horizon and numerical simulations confirm the conjecture that this creates stable relativistic jets, at least when the accretion disk is thick in the vertical direction~\cite{2012MNRAS.423.3083M} (Fig.~\ref{fig:mck}). 

The more complex electrodynamics of the ergosphere and relativistic innermost disk has to be included in these simulations. The magnetic flux responsible for the jets is believed to be advected inward from the outer radius of the accretion disk --- only a tiny fraction of the flux likely to thread the outer disk will suffice --- and to be constantly regenerated through ``magnetorotational'' instabilities~\cite{1998RvMP...70....1B}. They will exert a torque which will inevitably be dissipative as it drives gas inward. The gas in the disk will be maintained in a thermal state that balances heat generation, transport and radiative loss. The details of the many physical processes involved and their reconciliation with the observations is still a work in progress. What actually happens is thought to be mostly controlled by the mass supply to the disk in units of $L_E/c^2$. When this is high, the inner disk will be radiation-dominated and the photons will be trapped. They can either be advected into the hole or, more likely in AGN, be carried off in a wind. When the mass supply is low, the problem is that the plasma does not cool on an inflow timescale and a thick disk or torus is likely to form, supported by the pressure of mildly relativistic protons. Again, the failure to radiate the released binding energy is likely to result in a strong wind that caries away most of the energy instead. This wind may contribute to the transverse confinement of the jets. When the mass supply is intermediate, the accreted gas can radiate and cool and the disk will, consequently, be thin and there is no need for a wind.

In this interpretation, the relativistic jet, which is the source of the TeV photons, which HAWC will observe, (as well as the radio and optical synchrotron emission) is ultimately powered by the spin of the black hole.  By contrast, the quasi-thermal radiation, which is most noticeable in quasar infrared though ultraviolet light, is derived from the binding energy of the orbiting gas in the accretion disk. The all-important X-rays can originate from both the jet as well as a hot corona above the accretion disk. 

All of this may map onto the taxonomy and evolution of AGN. Many galactic nuclei may harbor black holes that accrete at the Eddington rate and shine with the Eddington luminosity while they are spun up by the angular momentum of the accreted gas as it plunges towards the event horizon from the minimum stable circular orbit. After this phase is over, the mass supply onto a more massive hole dwindles and the nuclear activity is fueled mainly by the spin of the hole. When the spin axis is directed towards us, we say that FSRQ evolve into HBL BLL! 

\section{Relativistic Jets}
\label{sec:jet}
Blazars are likely to dominate the HAWC source catalog. However, despite well-executed observational campaigns and discoveries, and great advances in simulation, their jets remain enigmatic. We are still debating many of the issues that were identified 40 years ago, specifically; hole vs disk, ions vs positrons, electromagnetic vs particle dominance, coherent vs synchrotron radio emission, near vs far $\gamma$-ray emission and so on. AGN jets are highly inhomogeneous and we observe them as they expand, over $\sim10^{10}$ in radius in the most extreme cases. Understanding them may depend upon frequency, angular resolution and epoch. In addition, relativistic beaming can be a very powerful amplifier of the emission from an otherwise insignificant part of the flow simply because it is temporarily moving with high speed in our direction. A further complication is that this speed is likely to differ from the speed of a feature observed using VLBI. The ``one zone'' models which are still a feature of many observational papers, are almost surely oversimplified. 

AGN jets are  probably created as electromagnetically-dominated outflows in which the magnetic field is wound up in a helix of increasing pitch angle propagating away from the black hole.  The velocity of the reference frame in which the electric field vanishes, presuming that the Lorentz invariants $B^2-E^2>0$, $\vec E\cdot\vec B=0$  will quickly become relativistic with $B>E$ and directed along the jet. This can be identified with a fluid velocity and the rest frame of the fluid as the comoving frame. The plasma that is needed to supply the electrical current and the space charge must be continuously created, in the magnetosphere above the black hole, either through pair production or through cross-field transport from the accretion disk. Its inertia is likely to be extremely small and the jet power is predominantly Poynting flux at this point. In other words, the ratio of the electromagnetic energy flux to the particle energy flux, $\sigma$, is large. The total axial current flowing along the jet is $I\sim(L_j/10^{45}{\rm erg s}^{-1})^{1/2}$~EA, where $L_j$ is the jet power. This current must return to the hole/disk when the outflow is no longer mostly electromagnetic. The overall current flow may be either dipolar or quadrupolar and this is observationally distinguishable through Faraday rotation studies. The associated potential difference across the jet is $\sim100(I/{\rm EA})$~EV (where the ``E'' stands for Exa or $10^{18}$). Now, we expect that plasma from around the jet will become entrained into the jet and that $\sigma$ will decrease with distance along the jet. From an electromagnetic point of view, the current will cross the jet. The current that flows perpendicular to the magnetic field, which we call ``industrious'' will do reversible work on the plasma; the current that flows along the electric field, which we call ``prodigal'' will lead to dissipation and entropy production. We expect that the momentum  flux will eventually become mechanical. When the jet power is weak relative to the luminosity of the disk, as happens in a Seyfert galaxy, the photons which are scattered by the electrons and positrons contribute a sort of electrical resistance leading to further weakening of the jet and a source of X-ray photons that can illuminate the disk from above as inferred from observations of iron lines.

However, the details of all of these processes are poorly understood and it is to observations that we must turn to get a better description of AGN jets.

\section{Particle Acceleration Mechanisms}
\label{sec:pam}
AGN jets radiate nonthermally and this requires that relativistic electrons (and positrons) be accelerated freely. Several mechanisms have been discussed and it is easy to imagine that they all play a role. The commonest proposal is that the industrious currents accelerate the plasma to highly supersonic speed and that strong shocks form in the relativistic outflow. This may happen because an obstacle, such as a molecular cloud is intercepted, perhaps when a jet moves transversely. Alternatively a faster moving outflow may run into a slower flow and a nonlinear wave breaks to form an internal shock. Strong shock waves are strong creators of entropy and, as Coulomb collisions are woefully ineffectual, it is reasonable to see the particle acceleration as the entropy production. In the case of non-relativistic shocks, there is a simple mechanism, diffusive shock acceleration, which, in its simplest form, bounces test particles $O(v_{\rm particle}/v_{\rm shock})$ times across the shock front with each shock passage increasing the particle energy by a fractional amount $O(v_{\rm shock}/v_{\rm particle})$ and transmits a power law distribution function~\cite{2001JPhG...27.1589K}. Typically, where this is observed, in the interplanetary and interstellar media, the efficiency is high so that the high energy particles have to be considered as part of the shock and magnetic field is also amplified. Put another way, a significant fraction of the large scale electrical current is carried by the high energy particles. Despite its obvious attraction, diffusive shock acceleration has at least three drawbacks in relativistic jets. The shocks are relativistic and the process is so strongly modified that it is not clear that it operates like this. In addition, if we are correct in our inference that the major dissipation happens initially in a highly magnetized flow, then the shocks are only weak and poor accelerators anyway. (What is commonly assumed is that jets start off as high $\sigma$, become low $\sigma$ and then regenerate magnetic field at a strong shock front. This is unusual.) The third problem is that strong shocks are likely to be separated by much more than an X-ray-emitting electron cooling length which is inconsistent  with X-ray observations of jets like that in M87. 

An even older mechanism derives from the observation that astrophysical jets, like their aeronautical counterparts, are likely to be very noisy. Random wave modes will propagate in the comoving frame and can accelerate charged particles usually though second order stochastic processes. However, stochastic acceleration is typically rather slow in a high sigma environment~\cite{1973A&A....26..161B}. However, it must be going on. 

As we have emphasized, the main {\it desideratum} of electromagnetic jet outflows is to convert large scale electromagnetic energy into particle energy. This has stimulated a renewed interest in magnetic reconnection. During magnetic reconnection magnetic field lines ``change partners''  by passing through a small region surrounding ``X-points'' where the gradients steepen so much that there is sufficient resistivity to allow the magnetic field lines to move through the plasma and to break flux freezing. Reconnection is observed to occur in laboratory and space plasmas and it leads to a release of energy as the change of topology allows the magnetic flux tubes to find a lower energy state. This can happen explosively, as in a solar flare. Reconnection is generally accompanied by particle acceleration but this is not generally a very efficient process when the speeds are non-relativistic. However all of this could change in a relativistic plasma and simulations show relatively efficient acceleration~\cite{2013ApJ...770..147C}. However, the general requirement of having all the flux pass through a small region at an X-point implies that the process is quite slow. As with stochastic acceleration, magnetic reconnection is almost certainly occurring and it could be responsible for most of the dissipation in jets. 

\section{The TeV Challenge}
\label{sec:tev}
TeV blazars therefore present three linked challenges, to accelerate electrons to very high energy in the presence of strong radiative loss, to vary on observed timescales that are short compared with the black hole sizes and to convert large volumes of predominantly electromagnetic energy into this channel with high efficiency. These features are similar to those that characterize observations of other high energy sources. This behavior is not confined to blazars. The Crab Nebula, which is $\sim10$ lt yr across, exhibits dramatic flares in the flux of $\sim300$~MeV photons about once a year, lasting for a few days and with variations on timescales as short as a few hours \cite{2014RPPh...77f6901B}. Although the photon energies are quite low, they are thought to be emitted as synchrotron radiation by electrons with energy $\sim5$~PeV. If so, this requires the accelerating electric field to be locally $\sim5$ times larger than the magnetic field. Of course most of the electromagnetic power generated by the Crab pulsar goes into the steady acceleration of lower energy the $\sim$~TeV electrons and positrons that emit near UV synchrotron radiation in a $\sim0.3$~mG magnetic field that represents most of the bolometric luminosity of the nebula. GRBs also exhibit rapid variability on timescales $\sim10$~ms. They also emit $\gamma$-rays with energies above $\sim100$~GeV which has been used to conclude that $\Gamma>1000$ (M\'esz\'aros, these proceedings). All three types of source are thought to involve the rapid dissipation of electromagnetic energy. However, even in the Crab Nebula, the formal potential differences that can be tapped, in principle, is sufficient to account for the particle acceleration. The challenge is to access it quickly.

\section{Magnetoluminescence, Electromagnetic Detonation and Gamma-Ray Scintillation}
\label{sec:mag}
\begin{figure}[t!]
        \includegraphics[scale=0.30]{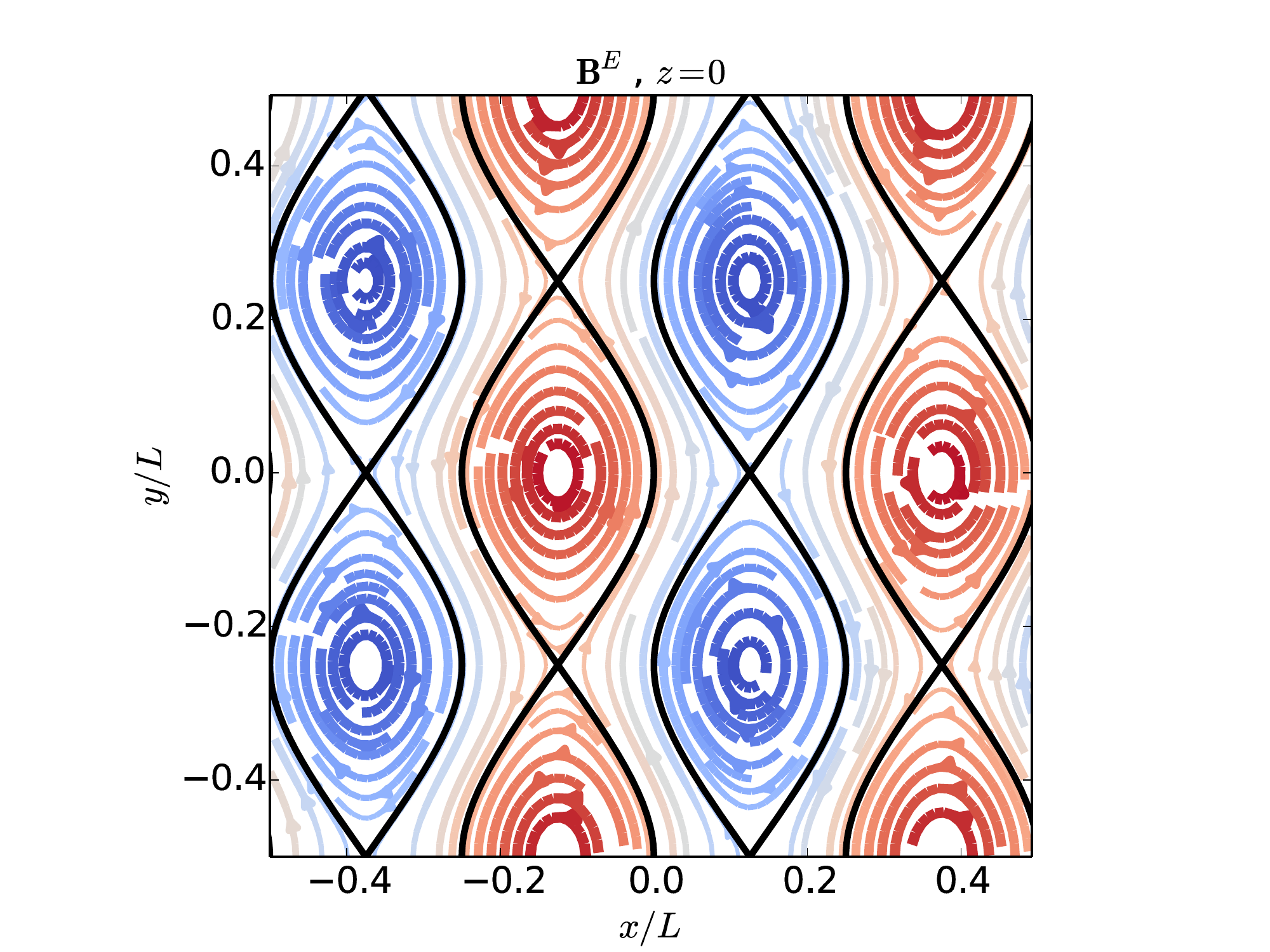}
        \includegraphics[scale=0.30]{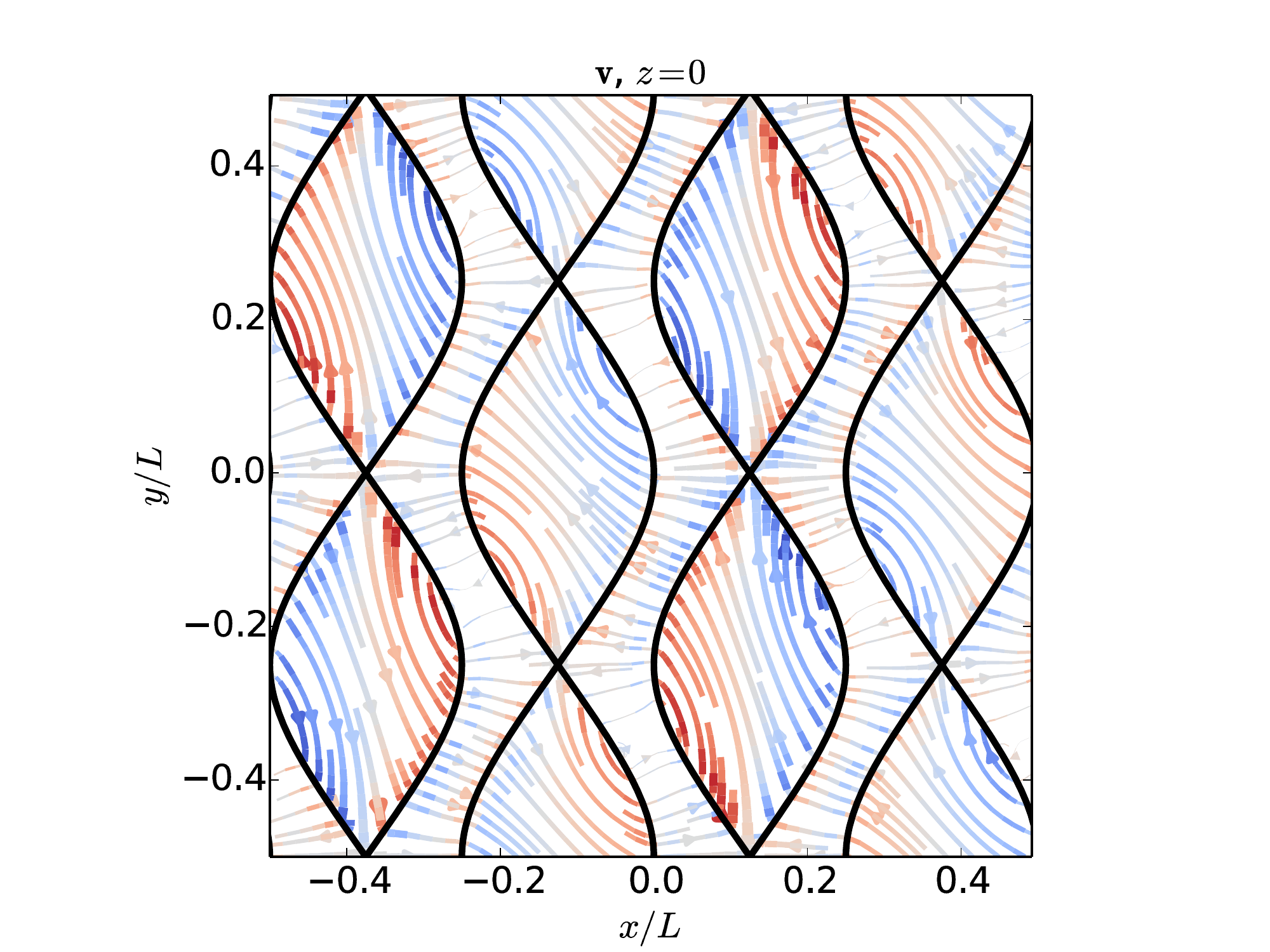}
        \caption{An example of an unstable Taylor state~\cite{2015arXiv150304793E}. The equilibrium magnetic field shown on the right (a so called "ABC" field) is unstable to the displacement given by the velocity field on the left.  Both plots show streamlines in the $z=0$ plane where the color represents the perpendicular vector component (with red indicating a component out of the page and blue indicating into the page) and the thickness of the line indicates the vector magnitude.  The black lines mark the separatrices in the equilibrium magnetic field.  In FFE and RMHD simulations, this instability acts to bring together like magnetic vortices, liberating magnetic energy on dynamical timescales and taking the field configuration to a longer wavelength state.\label{fig:rmhd}}
\end{figure}

We would now like to summarize some recent ideas which may help to address the TeV challenge as well as the larger task of accounting for the steady acceleration of lower energy particles in relativistic sources. The basic idea is that the tightly wound magnetic field in the inner jet is subject to dynamical instability~\cite{1998ApJ...493..291B} which is ultimately the cause of rapid $\gamma$-ray variability. Two formalisms can be invoked to describe a relativistic plasma where the energy density is dominated by electromagnetic field ~\cite{2002luml.conf..381B}. In Force-Free Electrodynamics (FFE), the inertia of the plasma is ignored and the charge density $\rho$ and current $\vec j$ satisfy $\rho\vec E+\vec j\times\vec B=0$. In Relativistic MHD (RMHD) the plasma is described as a relativistic fluid with a 4-velocity, energy density and pressure. In the limit of electromagnetic dominance and as long as $E<B$, these two approaches give similar results. The investigation that has been carried out is quite stylized but highly instructive~\cite{2015arXiv150304793E}. There are simple, three dimensional, force-free equilibrium solutions --- called generically Beltrami fields or Taylor states and, more specifically,  ``ABC'' solutions --- within a cubical box with periodic boundary conditions~(Fig.~\ref{fig:rmhd}). There are two relevant, global quantities that characterize these equilibria, the energy and the helicity which is expressed as $\int dV\vec A\cdot\vec B$ and which is a measure of the linkage of the magnetic flux. It is found that, contrary to earlier reports, equilibria are generically unstable if there is a lower energy state with the same helicity. The initial evolution is exponential but can be followed by intervals when the growth is much smaller before picking up again. Under ideal conditions, both the energy and the helicity are conserved.  However, the electric field can eventually grow to equal the magnetic field in strength. At this point energy, though not helicity, may be rapidly lost. In words, ropes of magnetic flux untangle but do not sever each other and their overall topology is preserved. How and in what form the energy is released is indeterminate under continuum mechanics but it is tempting to speculate that it can be used in impulsively accelerating particles up to their radiation reaction limit where $e\vec E\cdot\vec v$ balances the synchrotron or Compton radiative loss, presumably in the form of $\gamma$-rays.    In the simulations, this dissipation appears to be confined to a thin layer where $E\sim B$.  

Despite the tradition of transforming into a comoving frame and back again, it is actually more useful to work just in the galaxy frame and consider the counter-streaming, undulatory motion of the electrons and positrons subject to an electromagnetic acceleration $\pm (e/m)(\vec E+\vec v\times\vec B)$ with components both along and perpendicular to the velocity. The parallel component, $\pm(e/m)\vec E\cdot\vec v$, increases the particle energy; the perpendicular component, of magnitude $(e/m)[v^2B_\perp^2+E_\perp^2-2\vec E\times\vec B\cdot\vec v]^{1/2}$, is responsible for synchrotron-like radiation and can be used to compute the spectrum.  It is a curious fact that if the electrons are accelerated to an energy where the synchrotron radiative loss balances the electrical acceleration, then the characteristic energy of the emitted photons $\sim\gamma^2B$ is $\sim\alpha^{-1}m_ec^2(E/B)\sim70(E/B)$~MeV, independent of the electron energy $\sim\gamma m_ec^2$ and the strength of the magnetic field. (In the case of the Crab Nebula flares, where energies up to $\sim 400$~MeV are observed, $E\sim5$~B and the cooling length has to be $\sim0.2$~Larmor radii. This demonstrates the particle acceleration challenge!) However, in blazars, the dominant emission is through inverse Compton scattering of ambient or local synchrotron radiation. This may be in the Klein-Nishina limit at the highest $\gamma$-ray energies observed. There are spectral indications that this is the case.

It is therefore conjectured that the spinning black hole and its disk create anti-parallel jets increasingly dominated by toroidal magnetic field which become dynamically unstable and evolve to make two dimensional, dissipative, current sheets that propagate through the jet accelerating pairs at the expense of the electromagnetic energy and radiating  $\gamma$-rays. We call this general process ``magnetoluminescence''. This is meant as an analog of sonoluminescence and, like this phenomenon, probably involves an implosion not an explosion. This leads to significantly greater amounts of energy being made available as the external medium performs radial work and re-energizes the dissipating and cooling volume.

Most discussions of current sheets to date have been in the context of reconnection. An alternative possibility is an electromagnetic detonation. It is supposed that an incoming high $\sigma$ flow is changed into an outgoing, lower $\sigma$ momentum and energy flux. Within the transition, positrons and electrons are subjected to an unbalanced electric field and accelerated in opposite directions losing energy though $\gamma$-ray emission, simultaneous with the acceleration, not consequent to it as in a conventional radiative shock front. In general, this type of structure will have a much larger volumetric rate of conversion of electromagnetic energy into $\gamma$-rays than reconnection,  as it sweeps through a large region at the speed of light. Reconnection has to wait for magnetic flux to flow through a small region. The electrons and positrons have equal density and move with the frame in which the electric field vanishes both ahead of and behind the front but are displaced transversely in the front as a current sheet. Simple models have been constructed and will be described elsewhere. Their stability to fluid and kinetic perturbations is interesting. Coherent radio emission is also a possibility which may be relevant to the high brightness, fast radio bursts.  ``Particle in Cell'' simulations, including radiation reaction,  are underway to try to explore these and other schemes for rapid dissipation of electromagnetic energy.

There have been various attempts to account for the rapid, $\gamma$-ray variability. Some have involved extremely large Doppler factors; others have invoked a large number of independent emitters, simultaneously beaming their emission  in random directions --- ``minijets''~\cite{2011MNRAS.413..333N}.  A somewhat different possibility is suggested by the proposed emission from rapidly moving surfaces. If the direction of beamed emission varies smoothly in two dimensions over the surface, then conditions are ideal for the formation of caustics, similar to those found in refractive scintillation at much longer wavelengths. The spikes of emission can be associated with catastrophes when pairs of beams appear and disappear causing twinkling, similar to what is observed when a point source of light is reflected many times from the surface of a lake. 

\section{Conclusion}
\label{sec:con}
Observations of blazars, and related sources at $\gamma$-ray energies have already turned up a large number of surprises. These challenge pre-existing interpretations based largely upon radio, optical and X-ray observations. There are still serious debates about where the $\gamma$-rays originate. The most natural explanation, given the challenge of explaining the particle acceleration occurring in very small spacetime volumes, is to locate the emission as close to the central black hole as possible, given the pair production opacity. However, the evidence that this is generally the case is not yet compelling and this is one area where HAWC, with its enormous field of view, can excel, especially at the highest photon energies even when CTA is fully operational. The closer the sources are to the black hole, the higher the proportion of electromagnetic energy in the jet and the easier it will be to account for the emission. The prospect of pinpointing the emission sites using high frequency VLBI is very exciting indeed.

TeV astronomy is a very young field and there are surely several more major discoveries to be made, especially with VHE neutrino astronomy, gravitational radiation astronomy coming on line while the capabilities for transient astronomy at optical and radio wavelengths are ballooning. The fondest hope is that HAWC will surprise us all in its first years of operation.

\section*{Acknowledgements}
RB thanks the HAWC collaboration for the opportunity to attend and address this meeting. This work was supported in part by the U.S. Department of Energy contract to SLAC no. DE-AC02-76SF00515, NSF grant AST 12-12195, as well as the Simons Foundation, the Miller Foundation and the Humboldt Foundation (RB). YY gratefully acknowledges support from a KIPAC Gregory and Mary Chabolla Fellowship awarded to Stanford University.


\end{document}